\documentclass[12pt]{iopart}

\usepackage{graphicx}
\eqnobysec

\begin{document}

\title[On the stability of covariant BSSN formulation]{On the stability of covariant BSSN formulation}

\author{Ryosuke Urakawa$^1$, Takuya Tsuchiya$^2$ and Gen Yoneda$^3$}
\address{$^1$ Waseda University Advanced Research Institute for Science and Engineering, Okubo, Shinjuku, Tokyo 169-8555, Japan}
\ead{m-wasedamvps@ruri.waseda.jp}
\address{$^2$ Center for Liberal Arts and Sciences, Hachinohe Institute of Technology, 88-1, Obiraki, Myo, Hachinohe, Aomori 031-0814, Japan}
\address{$^3$ Department of Mathematics, School of Fundamental Science and Engineering, Waseda University, Okubo, Shinjuku, Tokyo 169-8555, Japan}

\begin{abstract}
  In this study, we investigate the numerical stability of the
  covariant BSSN (cBSSN) formulation proposed by Brown.
  We calculate the constraint
  amplification factor (CAF), which is an eigenvalue of the coefficient matrix
  of the evolution equations of the constraints on the cBSSN
  formulation and on some adjusted formulations with constraints added
  to the evolution equations.
  The adjusted formulations have a higher numerical stability
  than the cBSSN formulation from the viewpoint of the CAF.
\end{abstract}
\vspace{2pc}
\noindent{\it Keywords}: numerical relativity, eigenvalue analysis, covariant form

\submitto{\CQG}

\maketitle
\section{Introduction}
Since the first direct detection of gravitational waves \cite{wave} by Advanced
LIGO and Advanced VIRGO, various countries have been working on gravitational wave
observatories.
Many detections \cite{wave2,wave3,wave4,wave5,wave6,wave7} have been
contributing to the
knowledge of the universe and have paved the way for multi-messenger astronomy.
Numerical relativity plays an important role in finding
  gravitational waves.
In fact, simulations based on numerical relativity are important for acquiring a theoretical understanding of actually observed gravitational wave sources (e.g., \cite{wave}).

Simulations must be sufficiently accurate to withstand long calculations.
However, some simulations of strong gravitational sources are unsuccessful owing to numerical instability.
The Baumgarte--Shapiro--Shibata--Nakamura (BSSN) formulation \cite{BSSN0,BSSN1,BSSN2}, Z4 formulation \cite{Z4_1, Z4_6,Z4_7,Z4_8, Alic-2012-PhysRevD}, 
generalized harmonic (GH) formulation \cite{
    Garfinkle-2002-PhysRevD,
    Pretorius-2005-PhysRevLett,
    Pretorius-2005-CQG}
are well known to improve numerical stability. The spectral or pseudo-spectral methods  \cite{spectral1, spectral2, spectral3, spectral4, spectral5, improvedBSSN2} are also widely used in numerical relativists. There are also several comprehensive books in numerical relativity (e.g., \cite{alcubierre2008introduction, baumgarte2010numerical, Gourgoulhon-2012}). 
The numerical stability or instability is discussed in some papers \cite{instabilityBSSN1, instabilityBSSN2}.
These methods have been successfully applied to many complex astrophysical models and have contributed to the analysis of actual phenomena.
In simulations, it is of fundamental importance to pay attention to the stability and accuracy to guarantee the correctness of numerical results. The nonlinearity of the Einstein equations makes it easier to break constraints and we cannot perform longer simulations. Therefore, we focus on investigating the stability of equations and on how to improve it. In this study, we examine the BSSN formulation. 

It is well known that the BSSN formulation is better than the ADM
  formulation in performing the long-term simulations of strong
  gravitational phenomena.
  However, one of the disadvantages of the BSSN formulation compared with the ADM
  formulation is its noncovariant form.
  Thus, if we consider useful numerical techniques in the BSSN formulation
  in Cartesian coordinates, it is possible to invalidate to the formulation
  in other coordinates.
  Brown proposed the covariant form of the
  covariant BSSN (cBSSN) formulation \cite{cBSSN1},
  which has been studied in \cite{Z4_1, Alcubierre-2011-GRG, cBSSN_2012, cBSSN2, cBSSN3, Akbarian-2015-PhysRevD, cBSSN4, cBSSN5, cBBSN_2020, PRD104_084065}.
On the other hand, for the Z4 formulation the covariant form has been studied in \cite{covariantZ4_1, covariantZ4_2} for example.
  The construction of the cBSSN formulation
  is briefly described as follows.
There are five types of dynamical variable in the BSSN formulation, $(\phi, \tilde{\gamma}_{ij}, K, \tilde{A}_{ij}, \tilde{\Gamma}^i)$.
  The keys to deriving the covariant formulation are the
  definitions of the variable $\phi$, which is related to the conformal factor, and
  the variable $\tilde{\Gamma}^i$, which is related to the connection coefficient.
  In general, for the BSSN formulation, the metric variable $\tilde{\gamma}_{ij}$ is defined as $\tilde{\gamma}_{ij}=e^{-4\phi}\gamma_{ij}$, where $\gamma_{ij}$ is the spatial metric on the time-constant hypersurface and the second-order tensor.
  If we set $\phi$ as a scalar, then $\tilde{\gamma}_{ij}$
  becomes a tensor.
  In the same manner, the traceless-part of the extrinsic curvature $\tilde{A}_{ij}$
  becomes a tensor if $\phi$ is a scalar.
  Also, the trace-part of the extrinsic curvature $K$ is originally a scalar.
  In short, if we set $\phi$ and $\tilde{\Gamma}^i$
  as a tensor, then the
  formulation becomes covariant.
  The construction of the formulation is described in detail in \Sref{cBSSN}.

  It is important to investigate the numerical stability of the
    cBSSN formulation because the BSSN formulation was constructed to perform the long-term simulations on the strong
    gravitational spacetime.
    We can evaluate the stability of the formulation by obtaining the eigenvalues
    of the coefficient matrix of the evolution equations of the constraints.
    We refer to the eigenvalue as the constraint amplification factor (CAF) and the
    evolution equation of constraint as the constraint propagation equation (CPE).
  Some of the authors have studied techniques of controlling
  numerical stability and proposed adjusted systems.
  The basic idea of an adjusted system is to add constraints to evolution equations to control the violation of constraints. In \cite{Z4_7, Z4_8, adjusted1, adjusted2, adjusted3, adjusted4, improvedBSSN1}, 
  the efficiencies for stability are reported.

  The purpose of this study is to investigate the stability of the cBSSN formulation with CAFs
    and propose a more stable formulation using the adjusted
    system.
  For this purpose, we construct CPEs in the cBSSN formulation,
   apply this formulation to the adjusted system, and investigate CAFs and their stability in different backgrounds.
  
The structure of this paper is as follows.
First, we introduce the cBSSN formulation and the covariant evolution
equations of the matter fields in \Sref{cBSSN}.
Then, we review the CAF and the adjusted system in \Sref{adjusted}.
Next, we construct CPEs in the cBSSN formulation in \Sref{propagation}.
Then, we apply the cBSSN formulation to the adjusted system in \Sref{examples}.
Finally, we summarize our work in \Sref{summary}.
In this paper, we set the speed of light $c$ and the gravitational constant $G$ as unity, namely, $c=G=1$.
Latin indices such as $(i,j,k,\ldots )$ denote spatial indices and run from 1 to $n$, and Greek indices such as $(\mu,\nu,\ldots )$ denote spacetime indices and run from 0 to $n$, where n is the number of dimensions.

%
%
\section{Review of covariant BSSN formulation}\label{cBSSN}
In this section, we introduce the cBSSN formulation based on \cite{cBSSN1}.
For the basic variables of the ADM formulation $(\gamma_{ij}, K_{ij})$ where $\gamma_{ij}$ is the $n$-dimensional metric and $K_{ij}$ is the extrinsic curvature, those of the cBSSN formulation are defined as follows. 
The variable related to the conformal factor is defined as
\begin{equation}
  \phi = \frac{1}{4n}\log\frac{\mathrm{det}(\gamma_{ij})}{\mathrm{det}(f_{ij})},
  \label{phiDef}
\end{equation}
where $f_{ij}$ is any second-order positive definite tensor, and det means the determinant.
$\phi$ behaves as a scalar under all spatial coordinate transformations.
We define the scalar $W=e^{-2\phi}$ as a dynamical variable instead of $\phi$.
This is because the stability when using $W$ is higher than that when using $\phi$.
In particular,  
near a puncture black hole, some of the results are reported (e.g., \cite{cBSSN4}).
The conformal metric is defined as $\tilde{\gamma}_{ij}=W^2\gamma_{ij}$.
The trace-part of the extrinsic curvature is $K=\gamma^{ij}K_{ij}$,
and the traceless-part is $\tilde{A}_{ij} = W^2(K_{ij}-(1/n)K\gamma_{ij})$.
Also, for any second-order nondegenerate tensor $h_{ij}$, and its inverse $h^{ij}$, we define the variables
\begin{equation}
  H^i{}_{jk}=\frac{1}{2}h^{im}(\partial_jh_{km} + \partial_kh_{jm}
  - \partial_mh_{jk}),
  \label{HDef}
\end{equation}
and
\begin{equation}
  \Delta^i{}_{jk} = \tilde{\Gamma}^i{}_{jk}-H^i{}_{jk},
\end{equation}
where $\tilde{\Gamma}^i{}_{jk}=(1/2)\tilde{\gamma}^{im}
  (\partial_j\tilde{\gamma}_{mk}
  + \partial_k\tilde{\gamma}_{mj}
  - \partial_m\tilde{\gamma}_{jk}
  )$.
  $\Delta^i{}_{jk}$ behaves as a third-order tensor.
  We add a vector $\bar{\Lambda}^i$ as a new basic variable that satisfies
  $\bar{\Lambda}^i=\Delta^i$ at the initial time,
  where $\Delta^i
  = \tilde{\gamma}^{jk}\Delta^i{}_{jk}$.
Thus, we adopt $(W, \tilde{\gamma}_{ij}, K, \tilde{A}_{ij}, \bar{\Lambda}^i)$
as the new basic variables.
All basic variables are covariant under all spatial coordinate
  transformations.
  In this study, we assume that the second-order tensors $f_{ij}$ and $h_{ij}$ are constant with time.
    The Ricci tensor $R_{ij}$  is defined as
  \begin{eqnarray}
    R_{ij}
    &=& \tilde{R}_{ij} + R^W{}_{ij},\\
    \tilde{R}_{ij}
    &=&-\frac{1}{2}(\tilde{\gamma}^{mn}\mathcal{D}_m\mathcal{D}_n
      \tilde{\gamma}_{ij})
      + \tilde{\gamma}_{m(i}(\mathcal{D}_{j)}\bar{\Lambda}^m)
      \nonumber \\
      &&+ \Delta^\ell\Delta_{(ij)\ell}
      + 2\Delta^{mn}{}_{(i}\Delta_{j)mn}
      + \Delta^{mn}{}_i\Delta_{mnj},
    \\
    R^W{}_{ij}&=&
    (n-2)W^{-1}(\tilde{D}_i \tilde{D}_j W)
    +(1-n)W^{-2}(\tilde{D}_k W)(\tilde{D}^k W)\tilde{\gamma}_{ij} \nonumber
    \\
    &&+W^{-1}(\tilde{D}^k \tilde{D}_k W)\tilde{\gamma}_{ij},
  \end{eqnarray}
  where $\mathcal{D}_i$ is the covariant derivative operator
  associated with $h_{ij}$ and
  $\tilde{D}_i$ is the covariant derivative operator associated
  with $\tilde{\gamma}_{ij}$.
Both $\tilde{R}_{ij}$ and $R^W{}_{ij}$ behave as
  second-order tensors.
From the above, the constraint equations are 
\begin{eqnarray}
{\mathcal H}
&=&
W^2\tilde{R}
+(n-n^2)(\tilde{D}^k W)(\tilde{D}_k W)
+(2n-2)W(\tilde{D}^k \tilde{D}_k W), \nonumber
\\
&&+{n-1\over n}K^2
- \tilde{A}_{ij}\tilde{A}^{ij}
-2\Lambda -16\pi \rho_H =0, \label{const_H}
\\
{\mathcal M}_i
&=&
\tilde{D}^j\tilde{A}_{ji}
-n W^{-1}(\tilde{D}^j W)\tilde{A}_{ji}
+{1-n\over n}(\tilde{D}_i K)
-8\pi J_i =0,
\label{const_M}
\\
{\mathcal G}^i&=&
\bar{\Lambda}^i-\tilde{\gamma}^{jk}\Delta^i{}_{jk} =0, \label{const_G}
\\
{\mathcal S}&=&
\frac{\mathrm{det}(\tilde{\gamma}_{ij})}{\mathrm{det}(f_{ij})}-1 =0, \label{const_S}
\\
{\mathcal A}&=&
\tilde{A}_{ij}\tilde \gamma^{ij} =0, \label{const_A}
\end{eqnarray}
where $\tilde{R}=\tilde{\gamma}^{ij}\tilde{R}_{ij}$, and
  $\Lambda$ is the cosmological constant.
  For the energy momentum tensor $T_{\mu\nu}$,
  we have $\rho_H=n^\mu n^\nu T_{\mu\nu}$ where $n^\mu$ is the unit normal on the
  hypersurface of time constant $\Sigma(t)$, and
  $J_i =P{}^\mu{}_in^\nu T_{\mu\nu}$ where $P^\mu{}_i$ is the projection
  to $\Sigma(t)$. Here, we remark that these equations are satisfied analytically, but not always satisfied numerically. 
  
  The evolution equations are
\begin{eqnarray}
\partial_t W &=&
 {1\over n}\alpha WK
-{1\over n}W(\tilde{D}_k\beta^k)
+\beta^k (\tilde{D}_kW),
\label{evow}
\\
\partial_t \tilde{\gamma}_{ij}&=&
-2\alpha \tilde{A}_{ij}
-{2\over n}(\tilde{D}_k\beta^k)\tilde{\gamma}_{ij}
+2\tilde{\gamma}_{k(i}(\tilde{D}_{j)}\beta^k),
\label{evotg}
\\
\partial_t K&=&
{1\over n}\alpha K^2
+ \alpha \tilde{A}_{ij}\tilde{A}^{ij}
-W^2(\tilde{D}^i\tilde{D}_i\alpha) \nonumber \\
&&+(n-2)W(\tilde{D}^i\alpha)(\tilde{D}_iW)
+\beta^i(\tilde{D}_i K)
\nonumber \\&&
+{2\over 1-n}\alpha \Lambda 
+{8\pi\over n-1}\alpha S
+{8\pi(n-2)\over n-1} \alpha \rho_H,
\label{evok}
\\
\partial_t \tilde{A}_{ij}&=&
\alpha W^2 R^{\mathrm{TF}}_{ij}
+\alpha K\tilde{A}_{ij}
-2\alpha\tilde{A}_{ik}\tilde{A}^k{}_{j}
-8\pi W^2 \alpha S^\mathrm{TF}_{ij} \nonumber \\
&&-2W\{(\tilde{D}_{(i}\alpha)(\tilde{D}_{j)}W)\}^\mathrm{TF}
-W^2(\tilde{D}_i\tilde{D}_j\alpha)^\mathrm{TF}
+(\tilde{D}_i\beta^k)\tilde{A}_{jk} \nonumber \\
&&+(\tilde{D}_j\beta^k)\tilde{A}_{ik}
+ \beta^k(\tilde{D}_k\tilde{A}_{ij})
-{2\over n}(\tilde{D}_k\beta^k) \tilde{A}_{ij},
\label{evoa}
\\
\partial_t \bar{\Lambda}^i&=&
-2(\tilde{D}_j\alpha)\tilde{A}^{ij}
  - 2n\alpha W^{-1}(\tilde{D}_jW)\tilde{A}^{ij}
  + 2\alpha \Delta^i{}_{mn}\tilde{A}^{mn} \nonumber \\
  &&- \frac{2(n-1)}{n}\alpha \tilde{\gamma}^{ij}(\tilde{D}_jK)
  - 16\pi \alpha W^2 J^i
  + \tilde{\gamma}^{mn}(\mathcal{D}_m\mathcal{D}_n\beta^i)
  \nonumber\\
  &&
  + \frac{n-2}{n}(\tilde{D}^i\tilde{D}_m\beta^m)
  + \frac{2}{n}\Delta^i (\tilde{D}_m\beta^m) \nonumber \\&&
  + (\tilde{D}_j\bar{\Lambda}^i)\beta^j
  - \bar{\Lambda}^j(\tilde{D}_j\beta^i),
\label{evol2}
\end{eqnarray}
where $S_{ij}=P^\mu{}_iP^\nu{}_jT_{\mu\nu}$, $S=W^2\tilde{\gamma}^{ij}S_{ij}$,
$\alpha$ is the lapse function, $\beta^i$ is the shift vector, and
${}^{\mathrm{TF}}$ indicates the trace-free part of a second-order tensor.

 To derive CPEs, we need the evolution equations of the matter
 fields $\rho_H$ and $J_i$.
 The evolution equations of $\rho_H$ and $J_i$
 with the cBSSN variable are
\begin{eqnarray}
\partial_t \rho_H&=&
-\alpha W^2(\tilde{D}^k J_k)
+(n-2)\alpha W (\tilde{D}^k W) J_k
+\alpha K \rho_H
\nonumber \\&&
+\alpha W^2\tilde{A}^{mn}S_{mn}
+{1\over n} \alpha KS
-2(\tilde{D}^k \alpha)W^2 J_k
\nonumber \\&&
+\beta^k (\tilde{D}_k\rho_H),
\\
\partial_t J_i&=&
-\alpha W^2(\tilde{D}^k S_{ki})
+(n-2)\alpha W(\tilde{D}^k W)S_{k i}
-\alpha W^{-1}(\tilde{D}_i W)S
\nonumber \\&&
+\alpha K J_i
- W^2(\tilde{D}^k \alpha) S_{ki}
-(\tilde{D}_i\alpha)\rho_H 
\nonumber \\&&
+\beta^k (\tilde{D}_k J_i)
+(\tilde{D}_i \beta^k)J_k,
\end{eqnarray}
respectively.

  We mention how to set
  $f_{ij}$ in (\ref{phiDef}) and $h_{ij}$ in (\ref{HDef}).
  From the viewpoint of constructing the cBSSN formulation,
  both $f_{ij}$ and $h_{ij}$ are arbitrary second-order tensors.
  Since the cBSSN formulation should be consistent with the BSSN formulation in
  Cartesian coordinates, we set
  \begin{equation}
    f_{ij}=h_{ij}=\delta_{ij},
  \end{equation}
  in Cartesian coordinates, 
  where $\delta_{ij}$ is the Kronecker delta.
  For example, for spherical coordinates with $n=3$,
  \begin{equation}
    f_{ij}=h_{ij} = \mathrm{diag}(1,r^2,r^2\sin^2\theta),
  \end{equation}
  by the coordinate transformation from the Cartesian coordinates, where diag indicates the diagonal components.
  Then, the determinant of $f_{ij}$ is
  \begin{equation}
    \mathrm{det}(f_{ij})
    =r^4\sin^2\theta,
  \end{equation}
  the components of $H^m{}_{ij}$ 
  become
  \begin{eqnarray}
    &&H^1{}_{22}=-r,\quad
    H^1{}_{33}=-r\sin^2\theta,\quad
    \nonumber \\&&
    H^2{}_{12}=H^2{}_{21}=\frac{1}{r},\quad
    H^2{}_{33}=-\frac{1}{2}\sin2\theta,\quad
    \nonumber \\&&
    H^3{}_{13}=H^3{}_{31}=\frac{1}{r},\quad
    H^3{}_{23}=H^3{}_{32}=\frac{1}{\tan\theta},
  \end{eqnarray}
  and the other components of $H^m{}_{ij}$ are zero.
  When $f_{ij}$ and $h_{ij}$ are adopted, the reference metric is an elegant
  approach (e.g., \cite{Gourgoulhon-2012, cBSSN1}), and the
  results are consistent.
    In this study, we mathematically derive the conditions of
    $f_{ij}$ and $h_{ij}$ in terms of the covariance.
%
\section{Review of CAF and adjusted system}\label{adjusted}
In this section, we review CAF and the adjusted system to construct a more stable system.
\subsection{CAF}
When the evolution system with constraints is integrated numerically,
  the preservation of constraints
  is necessary for successful simulations.
  In general, constraint errors increase with the progress of the evolution.
  On the other hand, at the end of a simulation, the constraint errors should be sufficiently
  small.
  For the evolution system with constraints, we
  predict the behaviors of the constraint errors from CPEs.
  CPEs for the BSSN formulation, but not for the cBSSN formulation
 are expressed in \cite{adjusted3,adjusted6}.
  Thus, we derive CPEs for the cBSSN formulation later.

The constraints must be preserved at any time; however, the constraints are not satisfied because of numerical errors. 
Therefore, it is desirable to analyze the stability of the system in
advance and we derive ways of increasing the stability such as by adjusting the system.
One of the most popular methods of investigating stability is eigenvalue analysis.
First, we suppose that $u^a(x^i, t)$ is a set of dynamical variables and their evolution equations are
\begin{equation}
\partial_t u^a = f^a(u^b, \ \tilde{D}_c u^b,\ \cdots),
\end{equation}
where $f^a$ is the smooth function, and the (first class) constraints $C^a$ are
\begin{equation}
C^a(u^b,\ \tilde{D}_c u^b,\ \cdots) = 0. \label{eq:first_class_constraints}
\end{equation}
Again we remark that constraints are zero analytically, but not zero numerically.
Next, we suppose that the evolution equations of (\ref{eq:first_class_constraints}) (called CPEs) are expressed as follows
\begin{equation}
\partial_t C^a = g^a(C^b,\ \tilde{D}_c C^b,\ \cdots),  \label{eq:CPEs_inCAF}
\end{equation}
where $g^a$ is a function such as $g^a=0$ when $C^a=0$.
For the cBSSN formulation, the CPEs can be expressed as the form of (\ref{eq:CPEs_inCAF}) because the Einstein equations are the first class of constraints \cite{Frittelli-1997-PhyRevD}.
Finally, we apply the Fourier transformation to CPEs for analyzing eigenvalues as follows
\begin{equation}
\partial_t \hat{C}^a = \hat{g}^a(\hat{C^b}) = M^a_{\ b}\hat{C}^b,
\end{equation}
where the symbol hat $\wedge$ denotes the Fourier transformed value. We call the eigenvalues of the coefficient matrix $M^a_{\ b}$ the CAFs.

  For the eigenvalue analysis for CPEs,
  we usually set a suitable background metric for numerical
  simulations.
  This is because the coefficients of CPEs
  are expressed by the gauge and the dynamical variables,
  which are not constant.
  Although we obtain the CAF, we do not even know its sign, and 
  it is difficult or impossible to evaluate the stability of the system in many cases.
  For this reason, we should set a suitable background metric for the initial 
  and boundary conditions.
  In addition, there are some derivative terms of constraints in CPEs, which must be 
  replaced with discrete terms to obtain CAFs.
    For the cBSSN formulation, CPEs are expressed
    in a covariant form.
    There are covariant derivative terms in the equations, and with 
    reference to the Fourier transformation,
    we replace them with the rule $\tilde{D}\rightarrow \mathrm{i}\vec{k}$,
    where $\mathrm{i}$ is the imaginary unit and $\vec{k}$ is the
    single wave vector in local inertial coordinates.
If the real part of the CAF is negative, simulations are more stable than when the CAF has a positive real part because the violations of constraints are expected to decay. 

The CAFs in the Minkowski background ensure the stability of the
  simulations if the numerical conditions are near flat spacetime.
  However, most of the numerical conditions are far from flat spacetime
  because the main targets of numerical relativity are complex
  phenomena such as black holes.
  Thus, if we can perform stable and accurate simulations,
  we will be able to investigate the CAFs in the adapted spacetime for the simulations.

\subsection{Adjusted system}
\label{subsec:adjusyedSystem}
Next, we review the adjusted system.
For the evolution equations with constraints such as the cBSSN formulation, if we add the terms consisting of the constraints 
to the evolution equations, the system is equivalent in mathematically.
Moreover, adding some extra terms to the evolution equations is expected to change
  some of the constraints.
If the real parts of CAFs become smaller upon the modification of the
  evolution equations, the violations of constraints decay and the simulations become more stable and robust.
For the evolution equation of the dynamical system with constraints,
the evolution equation after modification by the
 adjusted system is expressed as
\begin{eqnarray}
\partial_t u^a = [\mathrm{original \ terms}] +
\kappa h^a(C^b, \tilde{D}_c C^b,\dots),
\label{eq:modfiedEq}
\end{eqnarray}
where $u^a$ is the dynamical variable, $C^a$ is the constraint,
and $h^a$ is a function with the constraints such that $h^a=0$ when $C^a=0$.
Note that the adjusted term $h^a$ includes covariant derivative terms of the constraints instead of partial derivative terms.
This is because to maintain the covariance of the evolution equations.
The constant $\kappa$ is the damping parameter.
For the evolution system with constraints, a modification such as
  (\ref{eq:modfiedEq}) is an equivalent transformation to the original system because constraints are zero analytically.
  Thus, the solutions of the system do not change analytically.
  Note that a better sign of $\kappa$ is generally obtained by calculating CAFs.
  However, we generally investigate which value is optimal when performing a numerical calculation.
\section{Constraint propagation equations of covariant BSSN formulation}\label{propagation}
In a previous study \cite{adjusted3}, CPEs of the BSSN formulation were introduced; however, they were not covariant and were only applied in the case of $n=3$. 
Here, we derive CPEs of the cBSSN formulation for any number of dimensions.
The equations are
\begin{eqnarray}
  \partial_t \mathcal{H}&=&
\frac{2}{n}\alpha K\mathcal{H}
-2(\tilde{D}^j\alpha)W^2 \mathcal{M}_j
+(2n-4)\alpha W(\tilde{D}^jW)\mathcal{M}_j
\nonumber \\&&
-2\alpha W^2\Delta^m{}_{m}{}^j\mathcal{M}_j
- 2\alpha\tilde{A}^{mn}W^2 (\Delta_{mn k}\mathcal{G}^k)^\mathrm{TF}
\nonumber \\&&
+\frac{2}{n}\alpha K W^2\Delta^l{}_{lk}\mathcal{G}^k
-\frac{2}{n}\alpha K W^2(\tilde{D}_k\mathcal{G}^k)
\nonumber \\&&
+ 2\alpha \tilde{A}^{mn}W^2(\tilde{D}_m\mathcal{G}_n)^\mathrm{TF}
+\frac{2n-2}{n}(\tilde{D}^k\tilde{D}_k\alpha)W^2\mathcal{A}
\nonumber \\&&
+\left(-2n+2-\frac{4}{n}\right)(\tilde{D}_k \alpha)W(\tilde{D}^kW) \mathcal{A}
+ \frac{2}{n}\alpha R \mathcal{A}
\nonumber \\&&
- \frac{2}{n}\alpha W^2(\tilde{D}_k\mathcal{G}^k) \mathcal{A}
+ \frac{2}{n}\alpha W^2\Delta^l{}_{lk}\mathcal{G}^k  \mathcal{A}
\nonumber \\&&
- \frac{16}{n}\pi  \alpha S \mathcal{A}
+(2n-2)\alpha (\tilde{D}^kW)(\tilde{D}_kW)\mathcal{A}
\nonumber \\&&
-2\alpha W(\tilde{D}^k \tilde{D}_k W)\mathcal{A}
+4(\tilde{D}^k\alpha )W^2(\tilde{D}_k\mathcal{A})
\nonumber \\&&
+(2-2n)\alpha W(\tilde{D}^kW)(\tilde{D}_k\mathcal{A})
+2\alpha W^2(\tilde{D}^k\tilde{D}_k\mathcal{A}) + \beta^k (D_k\mathcal{H}), \label{CP_H}
\\
\partial_t \mathcal{M}_i&=&
-\frac{1}{n}(\tilde{D}_i\alpha)\mathcal{H}
+\frac{n-2}{2n}\alpha (\tilde{D}_i\mathcal{H})
+\alpha K \mathcal{M}_i
\nonumber \\&&
+(2-n)\alpha W(\tilde{D}^m W) (\Delta_{mi k}\mathcal{G}^k)^\mathrm{TF}
\nonumber \\&&
+\frac{1}{n}(\tilde{D}_i\alpha)W^2\Delta^l{}_{lk}\mathcal{G}^k
+\frac{2-n}{n}\alpha W(\tilde{D}_iW)\Delta^l{}_{lk}\mathcal{G}^k
\nonumber \\&&
-\frac{1}{2}\alpha W^2(\tilde{D}_i\Delta^l{}_{lk})\mathcal{G}^k
+\alpha W^2 (\tilde{D}^m\Delta_{mi k})\mathcal{G}^k
\nonumber \\&&
+(\tilde{D}^m\alpha)W^2 (\Delta_{mi k}\mathcal{G}^k)^\mathrm{TF}
+(n-2)\alpha W(\tilde{D}^m W)(\tilde{D}_m\mathcal{G}_i)^\mathrm{TF}
\nonumber \\&&
-\frac{1}{n}(\tilde{D}_i\alpha)W^2(\tilde{D}_k\mathcal{G}^k)
+\frac{n-2}{n}\alpha W(\tilde{D}_iW)(\tilde{D}_k\mathcal{G}^k)
\nonumber \\&&
-\frac{1}{2}\alpha W^2\Delta^l{}_{lk}(\tilde{D}_i\mathcal{G}^k)
-(\tilde{D}^m\alpha) W^2(\tilde{D}_m\mathcal{G}_i)^\mathrm{TF}
\nonumber \\&&
+\alpha W^2 \Delta^m{}_{i k}(\tilde{D}_m\mathcal{G}^k)
+\frac{1}{2}\alpha W^2 (\tilde{D}_i\tilde{D}_k\mathcal{G}^k)
-\alpha W^2 (\tilde{D}^m \tilde{D}_m\mathcal{G}_i)
\nonumber \\&&
-(\tilde{D}_k \alpha) \tilde{A}_{i}{}^k\mathcal{A}
+\frac{1}{n}(\tilde{D}_i\tilde{D}_k\beta^k)\mathcal{A}
-\alpha \tilde{A}_{i}{}^k(\tilde{D}_k \mathcal{A})
\nonumber \\ &&
+(\tilde{D}_i \beta^k)\mathcal{M}_k
+\beta^k(\tilde{D}_k \mathcal{M}_i),  \label{CP_M}
\\
\partial_t \mathcal{G}^i&=&
2\alpha \tilde{\gamma}^{ji} \mathcal{M}_j 
-\alpha (\tilde{D}^i\mathcal{A})
+\frac{2}{n}(\tilde{D}_k\beta^k)\mathcal{G}^i,  \label{CP_G}
\\
\partial_t \mathcal{S}&=&
-2\alpha \mathcal{S}  \mathcal{A}
-2\alpha  \mathcal{A}, \label{CP_S}
\\
\partial_t \mathcal{A}&=&
\alpha K\mathcal{A} + \beta^k(\tilde{D}_k\mathcal{A}).  \label{CP_A}
\end{eqnarray}
If the initial condition satisfies the constraints, that is, $\mathcal{H}=0,\, \mathcal{M}_i=0,\, \mathcal{G}^i=0,\, \mathcal{S}=0$, and $\mathcal{A}=0$, the constraints are satisfied for all values of time 
because the time derivative of the constraints, (\ref{CP_H})--(\ref{CP_A}), are expressed in linear combinations of constraints, covariant derivatives of constraints, or their cross terms. Also, CPEs do not change under different matter fields
since the matter terms are not expressed explicitly in the above equations.

The CAF is unchanged by coordinate transformations that satisfy
  $\partial_t (\partial x^j / \partial x^{i'})=0$ because CPEs are covariant in coordinate transformations of space.
  This means that the numerical stability is unchanged for all
  nondynamical coordinate transformations.
  
In the next section, we investigate CAFs in some adjusted terms as examples and show their validity for simulations.

\section{Examples of CAFs in covariant BSSN formulation}\label{examples}
The Minkowski metric is the most basic solution of the Einstein
  equations.
  Furthermore, since this metric is simple, the analysis of the CAFs is simple and we approximately know the stability of simulations using the formulation.
  On the other hand, one of the main targets of numerical relativity
  is the analysis of black holes.
  The background metric is expressed as the Schwarzschild metric in the static case
  and as the Kerr metric in the rotational case.

In this section, we show some examples of adjusted systems
and the CAFs for the Minkowski, Schwarzschild, and Kerr backgrounds.
In the Minkowski background, we set the space dimension as $n\geq 3$.
On the other hand, for the Schwarzschild and Kerr
backgrounds, we set $n=3$.

\subsection{Minkowski background}
First, as a simple example, we show the CAF in the Minkowski background in the cBSSN formulation. 
In the Minkowski background, we set
$\alpha = 1$, $\beta^i = 0$, $\tilde{\gamma}_{ij} = \delta_{ij}$, $W=1$, $K=0$, $\tilde{A}_{ij}=0$, $\bar{\Lambda}^i=0$, and change $\tilde{D}_i$ to $\mathrm{i}k_i$ in (\ref{CP_H})--(\ref{CP_A}).
In addition, $f_{ij}=h_{ij}=\delta_{ij}$ because of the Cartesian coordinates in this case.
Then, the CPEs become
\begin{eqnarray}
\partial_t \hat{\mathcal{H}}&=&
-2 k^2 \hat{\mathcal{A}}, \label{total_H}
\\
\partial_t \hat{\mathcal{M}}_i&=&
\frac{n-2}{2n}\mathrm{i}k_i \hat{\mathcal{H}}
-\frac{1}{2}k_ik_j \hat{\mathcal{G}}^j
+k^2\delta_{ij}\hat{\mathcal{G}}^j, \label{total_M}
\\
\partial_t \hat{\mathcal{G}}^i&=&
2 \delta^{ij}\hat{\mathcal{M}}_j
-\mathrm{i}k_j\delta^{ij}\hat{\mathcal{A}}, \label{total_G}
\\
\partial_t \hat{\mathcal{S}}&=&
-2 \hat{\mathcal{A}}, \label{total_S}
\\
\partial_t \hat{\mathcal{A}}&=&
0, \label{total_A}
\end{eqnarray}
where $k=|\vec{k}|$ and $k_i$ is the component of $\vec{k}$, namely,
  $\vec{k}=(k_1,k_2,k_3,\cdots)$.
Therefore, the CAFs in the Minkowski background in the cBSSN formulation are
\begin{eqnarray}
\mbox{CAFs}=\left( 0, 0, 0, \pm k, \pm \sqrt{2}k (\times (n-1))\right) , \label{caf1}
\label{CAFs_Min}
\end{eqnarray}
where $\times$ denotes the number of iterations. We consider that the CPEs are $2n+3$ equations in total; thus, the number of CAFs is $2n+3$.
There are three zero values, $n$ positive real values, and $n$ negative real values.
That is, this system contains modes with growing numerical error due to the existence of positive CAFs, 
and modes with decaying errors due to the existence of negative CAFs.
This means that the stability of the cBSSN
  formulation in the Minkowski background is low because of the existences of the positive CAFs.
According to \cite{adjusted3}, the CAFs of the BSSN formulation in the case of $n=3$ in the
  Minkowski background are
  $(0,0,0,\pm\mathrm{i}k(\times6))$.
  Since the CAFs of the BSSN formulation do not include any positive real values,
  the cBSSN formulation is less stable than the BSSN formulation.
However, if we add the adjusted terms
  to the evolution equations of the cBSSN formulation so that the modes with growing error disappear, 
  then the adjusted system is more stable than the BSSN formulation.

The equations for, 
$\partial_t \hat{\mathcal{H}}$ (\ref{total_H}), $\partial_t\hat{\mathcal{G}}^i$ (\ref{total_G}), and $\partial_t \hat{\mathcal{S}}$ (\ref{total_S}) contain $\hat{\mathcal{A}}$ on the right-hand side.
If $\partial_t\tilde{\gamma}_{ij}$ or $\partial_t \tilde{A}_{ij}$
  is modified, the CAFs are expected to be affected.
Thus, in the Minkowski background, we adjust (\ref{evotg}).
The adjusted terms are essentially composed of the product of the constant $\kappa$, the lapse function or the shift vector (or its derivative), and the constraint (or its derivative). 
The effect of the adjusted terms in the evolution equations of
  the dynamical variables on CPEs
  is seen in the variation of the constraints, which is explicitly shown in Appendix.
  We suggest two adjustments for the dimension $n=3$.

\subsubsection{Case I}
\begin{figure}[htbp]
\centering
\includegraphics[width=10cm]{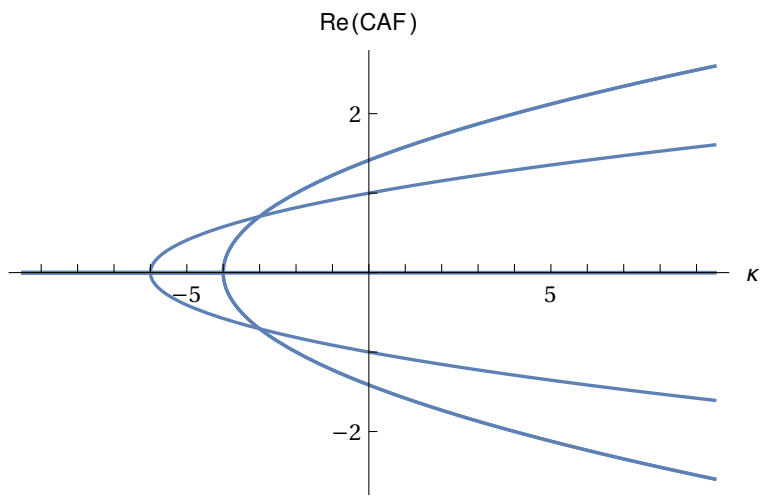}
\caption{Real parts of the CAFs in the Minkowski background obtained using (\ref{evow})--(\ref{evol2}) and (\ref{mod3}) with the wave vector set to $\vec{k}=(1, 0, 0)$. The horizontal axis indicates the damping parameter $\kappa$ and the vertical axis indicates the real part of each CAF (Re(CAF)). There are nine curves in total, with some curves overlapping.  All curves are completely symmetric with respect to the horizontal axis.}
\label{fig:CAF1}
\end{figure}

We add the adjusted term to (\ref{evotg}) as
\begin{eqnarray}
\partial_t \tilde{\gamma}_{ij} = \mathrm{[original\ terms]} + \kappa \alpha \tilde{D}_{(i}\mathcal{M}_{j)}.
\label{mod3}
\end{eqnarray}
The real parts of nine CAFs in the Minkowski background are shown in \Fref{fig:CAF1}.
This graph provides some information on the numerical stability with respect to each damping parameter $\kappa$. For example, in the case of $\kappa = 0$, the CAFs are  $0$, $\pm k,$ and $\pm\sqrt{2}k$, as can be seen from (\ref{CAFs_Min}). In the case of $\kappa \leq -6$, the system is more stable than that in the case of $\kappa = 0$ because the positive real part of the CAFs disappears.
In contrast, if $\kappa$ is set to more than zero, this system is less stable than that in the case of $\kappa = 0$ because some positive real parts of the CAFs become larger. In conclusion, $\kappa = -6$ is the best value for tuning the damping parameter according to this graph, but we consider that the best value of $\kappa$ determined from this graph is not necessarily the same as that in real simulations.
It is important to know the sign of $\kappa$.
The discrepancy between the results of analyzing the CAFs and
  the actual simulations is mainly caused by the lack of information on the discretization.
  As mentioned in \Sref{subsec:adjusyedSystem}, the damping parameter
  $\kappa$ depends on the numerical grid.
  This is because the numerical errors are partly caused by the
  truncation errors in the evolution equations; thus, the magnitudes of the
  modified terms depend on the numerical grid.
  On the other hand, we obtain the sign of $\kappa$ from the CAFs
  from the analytical result at a continuous level.
  Therefore, the sign of $\kappa$ is irrelevant to the discrete information
  such as the numerical grid.

In this case, the CAFs are explicitly expressed as
  \begin{eqnarray}
    \mathrm{CAFs}=\left(
    0,0,0,\pm \frac{k\sqrt{4+\kappa k^2}}{\sqrt{2}},
    \pm \frac{k\sqrt{4+\kappa k^2}}{\sqrt{2}},
    \pm \frac{k\sqrt{6+\kappa k^2}}{\sqrt{2}}
    \right).
    \label{eq:CAFCaseI}
  \end{eqnarray}
  If we set $\kappa k^2\leq-6$, there are three zeros and six imaginary
    numbers.
As described at the beginning of this section, the cBSSN formulation
is less stable than the BSSN formulation.
In this case, we conclude that the modified cBSSN formulation with the adjusted equation (\ref{mod3}) having a negative $\kappa$ is expected to have the same stability as the BSSN
formulation in actual simulations. 
  
  From (\ref{eq:CAFCaseI}), the negative and positive real
  parts of the CAFs appear simultaneously when we set $\kappa>-6/k^2$.
  On the other hand, when $\kappa<-6/k^2$, the CAFs only include the imaginary
  part.
  Thus, the stability of the cBSSN formulation using (\ref{mod3}) is low.
  We remark that the adjusted term appears to be effective for the stability if the system breaks time-reversal symmetry \cite{adjusted3}. Here, time-reversal symmetry refers to the parity of a variable not changing if the time derivative is set to the opposite direction. Therefore, we consider adjusted terms that break time-reversal symmetry in the next case.

\subsubsection{Case II}
Next, we add the adjusted term to (\ref{evotg}) as
\begin{eqnarray}
\partial_t \tilde{\gamma}_{ij} =  \mathrm{[original\ terms]} + \kappa \alpha \tilde{D}_{(i}\mathcal{G}_{j)}.
\label{mod4}
\end{eqnarray}
The time-reversal parity of the left-hand side of (\ref{mod4}) changes
  because $\tilde{\gamma}_{ij}$ does not change with the time direction.
  On the other hand, the parity of the original terms on the right-hand side of
  (\ref{mod4}), namely, the right-hand side of (\ref{evotg}), changes with the time direction.
  However, the parity of the adjusted term of (\ref{mod4}), which is $\alpha
  \tilde{D}_{(i}\mathcal{G}_{j)}$, does not change with the time direction.
  Thus, the adjustment breaks the time-reversal parity and the stability appears
  to be improved by the adjustment.
\begin{figure}[htbp]
\centering
\includegraphics[width=10cm]{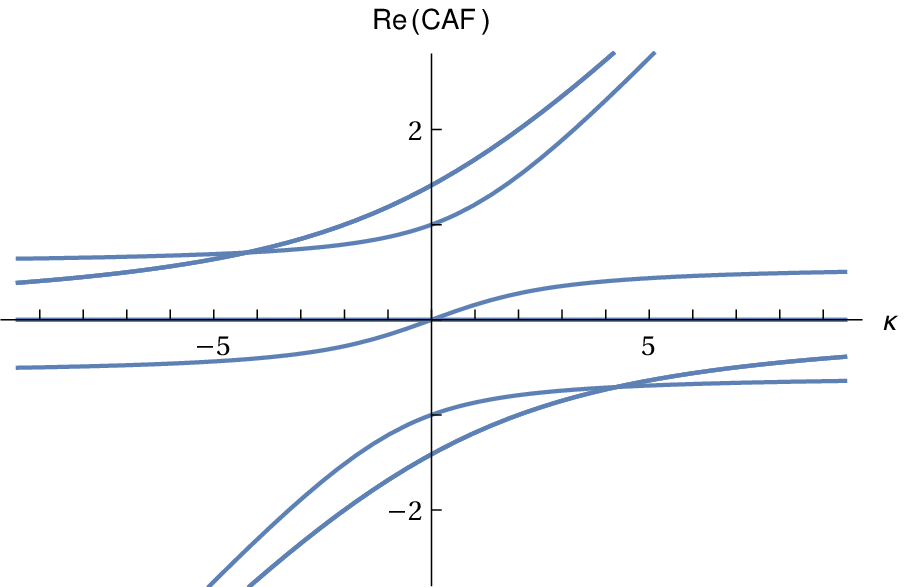}
\caption{As \Fref{fig:CAF1} but with the adjusted equations, i.e., 
   (\ref{evow})--(\ref{evol2}) and
    (\ref{mod4}).
 There are nine curves
  in total, with some curves overlapping.
  Note that the values of $\mathrm{Re}(\mathrm{CAF})$ at $\kappa=0$ are the same as those in \Fref{fig:CAF1}.
}
\label{fig:CAF2}
\end{figure}

\Fref{fig:CAF2} shows the CAFs with the same numerical settings as in \Fref{fig:CAF1} except for the adjusted term of $\tilde{\gamma}_{ij}$. As $\kappa$ increases, some positive real parts of the CAFs become larger and some negative real parts of the CAFs approach zero, and the opposite occurs with decreasing $\kappa$. 
Since all the real parts of the CAFs for $\kappa<0$ are smaller than those for $\kappa = 0$,
  the adjusted system using (\ref{mod4}) is more stable than the unadjusted system of the cBSSN formulation.

  Although covariance is obtained, the stability is lost
  when changing from the BSSN formulation to the cBSSN formulation.
  For the two adjusted cases
  expressed by (\ref{mod3}) and (\ref{mod4}),
  we show the possibility of constructing more stable systems for the cBSSN formulation.
  In Case I, we show the construction of a system with the same stability as the BSSN
  formulation.
  We also show the construction of a more stable system than the unadjusted
  cBSSN formulation in Case II.
  It is expected that more stable systems can be constructed using other adjustments
   such as multiple adjustments based on (\ref{mod3}) and (\ref{mod4}).

\subsection{Schwarzschild background}

Next, we consider the Schwarzschild background. 
The Schwarzschild metric is given as 
\begin{equation}
  ds^2 = -\left(1-\frac{2M}{r}\right)dt^2
  + \left( 1-\frac{2M}{r}\right)^{-1}dr^2+r^2d\theta^2+r^2\sin^2\theta d\varphi^2.
\end{equation}
In this background, we adopt polar coordinates and $f_{ij}=h_{ij}=\mathrm{diag}(1,r^2,r^2\sin^2\theta)$.
\begin{figure}[htbp]
\centering
\includegraphics[width=10cm]{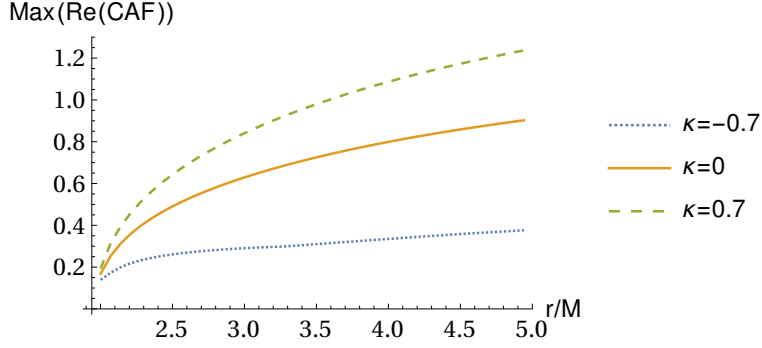}
\caption{Maximal real part of the CAFs in the
  Schwarzschild background obtained using (\ref{evow})--(\ref{evol2}) and
    (\ref{mod5}) with the wave vector set to $\vec{k}=(1, 1, 1)$. The horizontal axis indicates the radius divided by the mass $r/M$, and the vertical axis indicates the maximal real part of each CAF (Max(Re(CAF))).
  Since $r=2M$ is a singularity, these lines are drawn for $2.001\leq r/M\leq 5$.
  $\kappa$ is the damping parameter of (\ref{mod5}).
    The solid line show the result for $\kappa=0$,
    the dotted line shows that for $\kappa=-0.7$,
    and the dashed line shows that for $\kappa=0.7$.
    $\kappa=0$ is consistent with the case of the unadjusted cBSSN formulation.
}
\label{fig:CAF3_0}
\end{figure}
\Fref{fig:CAF3_0} shows the maximal real part of the CAFs in the Schwarzschild
background with the adjustment
\begin{eqnarray}
\partial_t \bar{\Lambda}^i
=  \mathrm{[original\ terms]} + 
\kappa \alpha
\left(W+\frac{2}{3}\right)^2\mathcal{M}^i.
\label{mod5}
\end{eqnarray}
In this figure, lines are drawn for $2.001\leq r/M\leq 5$ because of the singularity at $r=2M$.
The solid line in \Fref{fig:CAF3_0} shows the maximal real part of the CAFs
  with the unadjusted system of the cBSSN formulation in the Schwarzschild
  background.
The horizontal axis indicates the radius divided by the mass $r/M$.
  We see that the cBSSN formulation becomes stable from
  $r=5M$ to $r=2M$.
   On the other hand, in \cite{adjusted4},
  the ADM formulation becomes unstable from $r=5M$ to $r=2M$ in the
  Schwarzschild background.
For the adjusted cBSSN formulation with (\ref{mod5}),
  the formulation is more stable with the damping parameter $\kappa=
  -0.7$ than with $\kappa=0.7$.
  This means that negative values of $\kappa$ gives a higher stability than non-negative values of $\kappa$.

In this case, we comment on why the maximal real
  part of the CAFs becomes smaller toward the event horizon of $r=2M$.
  For the ADM formulation, the stability of the system decreases
  toward the event horizon from $r>2M$ \cite{adjusted4}.
  For the cBSSN formulation, the constraint propagation equations
  (\ref{CP_H})--(\ref{CP_A}) are represented by the dominant terms
  in the Schwarzschild background as
\begin{eqnarray}
  \partial_t \mathcal{H}&=&
-2(\tilde{D}^j\alpha)W^2 \mathcal{M}_j
+2\alpha W(\tilde{D}^jW)\mathcal{M}_j,
\label{CP_H_Sch}
\\
\partial_t \mathcal{M}_i&=&
\frac{1}{6}\alpha\tilde{D}_i\mathcal{H}
+\frac{1}{2}\alpha W^2 (\tilde{D}_i\tilde{D}_k\mathcal{G}^k)
-\alpha W^2 (\tilde{D}^m \tilde{D}_m\mathcal{G}_i),
\label{CP_M_Sch}
\\
\partial_t \mathcal{G}^i&=&
2\alpha \tilde{\gamma}^{ji} \mathcal{M}_j,
\label{CP_G_Sch}
\\
\partial_t \mathcal{S}&=& 0,
\label{CP_S_Sch}
\\
\partial_t \mathcal{A}&=& 0.
\label{CP_A_Sch}
\end{eqnarray}
Here, $K=0,\  \beta^i=0,$ and $\tilde{A}^{ij}=0$ in the Schwarzschild background, then $\mathcal{A}=0$ because $\partial_t\mathcal{A}=0$ and $\mathcal{S}=0$ because $\partial_t\mathcal{S}=0$ in the above expression.
  For the range of $2<r/M<5$,
  the magnitudes of $\alpha$ and $W$ approximately satisfy
  $|\alpha|<|W|<1$ and
  $|\alpha \tilde{\gamma}^{ij}|<|\alpha W(\tilde{D}^iW)|<|(\tilde{D}^i\alpha)W^2|<|\alpha W^2|$.
  Thus, the behavior of the maximal real part of CAFs almost
  follows that of $\alpha W^2$.
  Since the magnitudes of both $\alpha$ and $W$ decrease toward
  $r=2M$, the line in the case of $\kappa=0$  in \Fref{fig:CAF3_0}
  drops from $r>2M$ toward $r=2M$.

\subsection{Kerr background}
Finally, we consider the Kerr background.
The metric in Boyer--Lindquist coordinates is given as
\begin{eqnarray}
  ds^2 &=& -\left(1-\frac{2Mr}{\Sigma}\right)dt^2
    - \frac{4aMr\sin^2\theta}{\Sigma}dtd\varphi
    + \frac{\Sigma}{\Delta}dr^2
    \nonumber \\&&
    + \Sigma d\theta^2
    + \left(r^2+a^2+\frac{2a^2 Mr\sin^2\theta}{\Sigma}\right)\sin^2\theta
    d\varphi^2,
  \label{Kerr}
\end{eqnarray}
where $\Delta=r^2-2Mr+a^2$ and $\Sigma=r^2+a^2\cos^2\theta$.
One of the differences between the Schwarzschild metric and the Kerr metric
  is the existence of the shift term, i.e.,
  the cross term of $dtd\varphi$ in (\ref{Kerr}).
  Thus, the shift vector $\beta^i$ in the spacetime--decomposed component
  is not zero.
  Then, we add the adjusted terms with the $\beta^i$ component to the $\bar{\Lambda}^i$ evolution equation:
\begin{eqnarray}
  \partial_t \bar{\Lambda}^i &=&  \mathrm{[original\ terms]} + \kappa \{\alpha (W+W^2)\mathcal{M}^i
  \nonumber \\&&
  + c\beta^i
  (\tilde{D}_j\log(W^2+1))\bar{\Lambda}^j\mathcal{H}\},
\label{mod6}
\end{eqnarray}
where $c$ is a constant
  used to match the physical dimension.
  Here, we set $c=1/2$.
\begin{figure}[h]
  \begin{minipage}[b]{0.45\linewidth}
    \centering
    \includegraphics[keepaspectratio, scale=0.65]{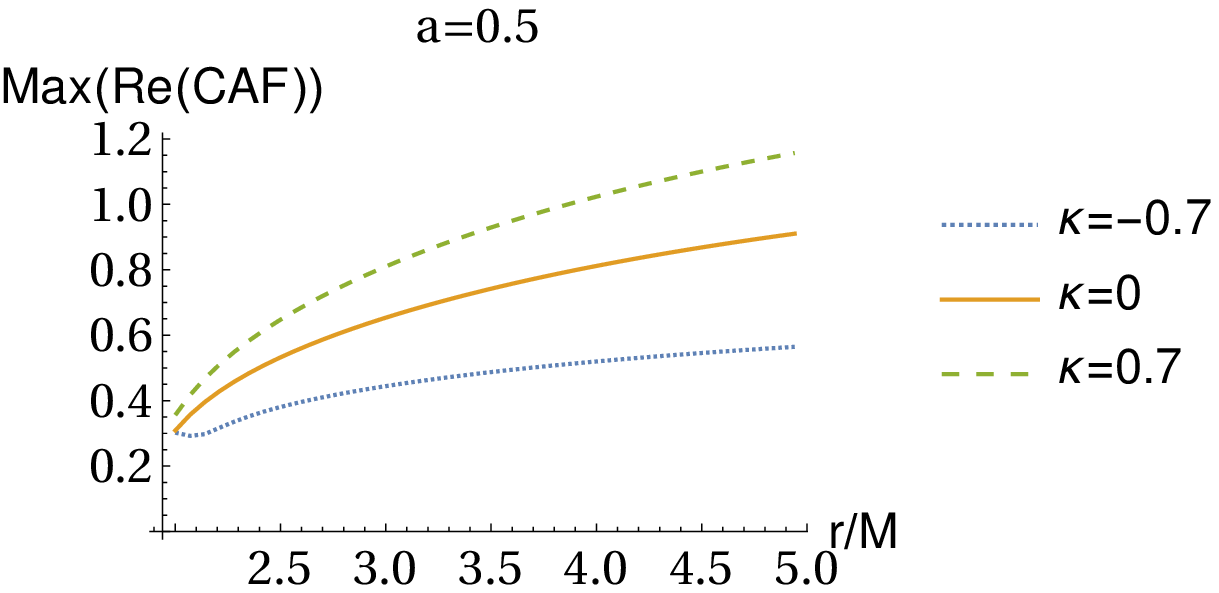}
  \end{minipage}
  \begin{minipage}[b]{0.45\linewidth}
    \centering
    \includegraphics[keepaspectratio, scale=0.65]{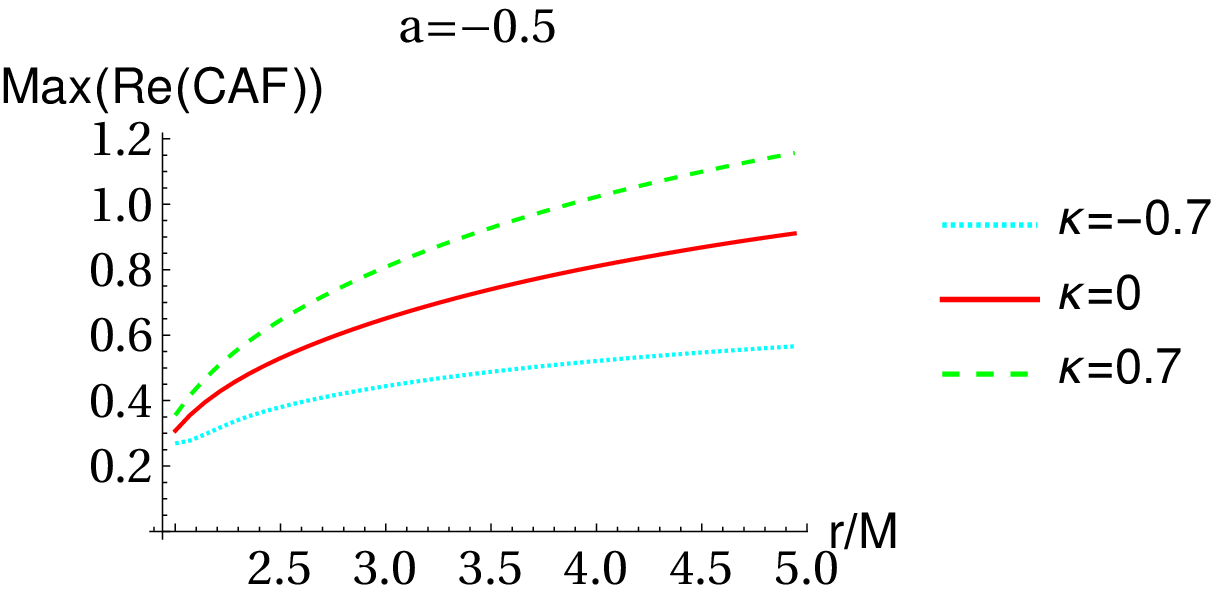}
  \end{minipage}
  \begin{minipage}[b]{0.9\linewidth}
  \centering
  \vspace{0.3cm}
    \includegraphics[keepaspectratio, scale=0.65]{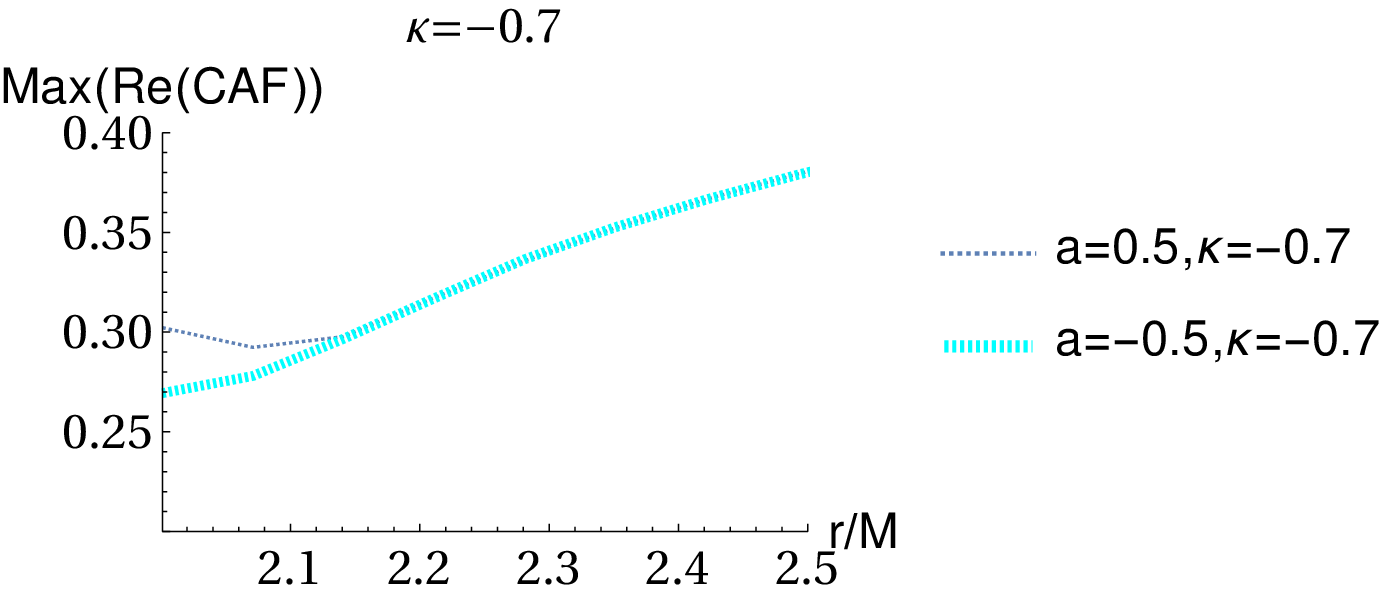}
  \end{minipage}
  \caption{
    As \Fref{fig:CAF3_0}
    except for
      the background metric and the adjusted equations.
      These graphs are drawn for the Kerr background metric with
      (\ref{evow})--(\ref{evol2}) and (\ref{mod6}).
      The upper left and right panel are drawn with the rotating parameter $a=0.5$
      and $a=-0.5$, respectively, in the range of $2.001\leq r/M\leq 5$.
      For upper two panels,
      the solid line represents the damping parameter $\kappa=0$,
      the dotted line represents $\kappa=-0.7$, and 
      the dashed line represents $\kappa=0.7$.
      For the lower panel, the thin dotted line represents $a=0.5$ and $\kappa =-0.7$, and the thick dotted line represents $a=-0.5$ and $\kappa =-0.7$.
      The lower panel is drawn with only two cases, the parameter $a=0.5,\ \kappa =-0.7$ and $a=-0.5,\ \kappa =-0.7$ in the range of $2.001\leq r/M\leq 2.5$. These two lines are slightly different in the range of $r/M < 2.2$. In the other range, they overlaps. The lines with  $\kappa =0.7$, and the ones with $\kappa =0$ in the upper panels overlap, respectively.
  }
  \label{fig:CAF4}
\end{figure}

\Fref{fig:CAF4} shows the maximal real part of CAFs in the Kerr background for the rotating parameter $a=0.5$ and $a=-0.5$, and as the damping parameter in
  (\ref{mod6}) $\kappa=-0.7, \, 0,$ and $0.7$.
By comparing the solid line in Figures \ref{fig:CAF3_0} and \ref{fig:CAF4}, we see that the
  maximal real part of the CAFs in the Kerr background is larger than
  that in the Schwarzschild background.
  This result indicates that the cBSSN formulation in the Kerr background is less stable than that in the
  Schwarzschild background.
In the cases of $\kappa = 0.7$ and
  $\kappa=0$ in the top panels, the system is less stable than that in the case of no
adjustment.
On the other hand, the system in the case of the negative damping parameter, $\kappa = -0.7$, has totally good effect.
From the lower panel in \Fref{fig:CAF4}, we see
  a slight difference in the dotted line ($\kappa=-0.7$) near $r=2M$.
  This difference seems to be caused by the adjusted term $c\beta^i(\tilde{D}_j\log (W^2+1))
  \bar{\Lambda}^j\mathcal{H}$ in (\ref{mod6}).
  This is because the sign of $\beta^i$ is consistent with that of the rotating parameter $a$.
  We conclude that the negative values of the damping parameter
are effective with the adjustment (\ref{mod6}).

  The main reason for the decrease in the maximal real part of the
    CAFs from $r>2M$ to $r=2M$ is the same as that for the Schwarzschild
    background.
    For the Kerr metric, some terms contain the rotating parameter
    $a$.
    Compared with the total mass $M$, the rotating parameter is small.
    Then, since the effect of the terms with the rotating parameter on the CAF
    is small,
    the behaviors are almost the same as in the Schwarzschild case.

\section{Summary}\label{summary}

We reviewed the construction of the covariant BSSN (cBSSN) formulation and the eigenvalue analysis of the evolution equations of constraints.
In addition, a method of forming a stable dynamical
system with constraints, referred to as the adjusted system, was shown.
Also, we introduced the covariant form of the constraint propagation
equations (CPEs), which are the evolution equations of constraints.

To study the stability of the cBSSN formulation, we calculated the constraint amplification factors (CAFs),
which are the eigenvalues of the coefficient matrix of CPEs,
in the Minkowski, Schwarzschild, and Kerr backgrounds.
We found that the cBSSN formulation is unstable, consistent with the original
BSSN formulation, but showed that it is possible to
improve the numerical stability so that a simulation becomes stable for some
  adjusted systems from the CAFs in the Minkowski background.
Next, for the Schwarzschild background, we proposed an adjusted formulation that is
 more stable than the unadjusted one with a negative damping parameter.
  In addition, we commented that the maximal real part of the CAFs becomes smaller
  from the outside of the event horizon, $r>2M$, to $r=2M$.
As the final example of the background, we showed the CAFs in the Kerr
  metric.
  We suggested an adjusted formulation and showed 
   that it was more stable than the unadjusted
  formulation with the negative damping parameter.
  We also investigated the effect of the rotating parameter $a$ and
  found that the stability deterioritates if $a\neq 0$.
  There is almost no difference for different signs of the rotating
  parameter.

  In this paper, we suggested some adjustments and calculated the CAFs.
    There are several reported cases where the stability of the CAFs coincides
    with that of the numerical calculations
    \cite{adjusted1,adjusted2,adjusted5,adjusted8};
    thus, it is expected that the numerical calculations are more stable when
    using the adjustments.
    This is under investigation and will be reported in the near future.
  
\appendix
\section*{Acknowledgements}
We thank referees for carefully reading and giving fruitful comments on our paper.
G.Y. and T.T. were partially supported by JSPS KAKENHI Grant Number
  20K03740.
  G.Y. was partially supported by a Waseda University Grant for Special
Research Projects 2021C-138. 
  T.T. was partially supported by JSPS KAKENHI Grant Number 21K03354,
  and a Grant for Basic Science Research Projects from
  The Sumitomo Foundation.


%
%
%
\appendix
\section*{Appendix}
\setcounter{section}{1}
We derive the total derivative of the constraint equations. These equations are very useful to investigate the difference of the constraint equations easily
if we change the constraint equations by adjustment of the evolution equations.

\begin{eqnarray}
\delta {\mathcal H}
&=&
\{
2W\tilde{R}
+(2n-2)(\tilde{D}^k \tilde{D}_k W)
\}
(\delta W)
+(2n-2n^2)(\tilde{D}^k W)(\tilde{D}_k\delta W)
\nonumber \\&&
+(2n-2)W(\tilde{D}^k\tilde{D}_k\delta W)
+\{
-W^2 \tilde{R}^{mn}
+W^2\tilde{D}^m \mathcal{G}^n
-W^2 \Delta^{mn}{}_k\mathcal{G}^k
\nonumber \\&&
+(n^2-n)(\tilde{D}^m W)(\tilde{D}^n W)
+ 2\tilde{A}^n{}_{j}\tilde{A}^{mj}
+(2-2n)W(\tilde{D}^m \tilde{D}^n W)
\nonumber \\&&
+W^2(\tilde{D}_k\Delta^{kmn})
-W^2\Delta^k{}_{kp}\Delta^{pmn}
\}
(\delta\tilde{\gamma}_{mn})
\nonumber \\&&
+\{
(2-2n)W(\tilde{D}^n W)\tilde{\gamma}^{lm}
+(n-1)W(\tilde{D}^l W)\tilde{\gamma}^{mn}
+W^2\Delta^{\ell mn}
\nonumber \\&&
+W^2\Delta^k{}_{k}{}^m\tilde{\gamma}^{\ell n}
-\frac{1}{2}W^2\Delta^k{}_{k}{}^\ell \tilde{\gamma}^{mn}
\}
(\tilde{D}_l \delta\tilde{\gamma}_{mn})
\nonumber \\&&
-\frac{1}{2}W^2\tilde{\gamma}^{\ell p}\tilde{\gamma}^{nm}(\tilde{D}_\ell\tilde{D}_p \delta\tilde{\gamma}_{mn})
+\frac{2n-2}{n}K(\delta K)
\nonumber \\&&
- 2\tilde{A}^{mn}(\delta\tilde{A}_{mn})
-W^2\Delta^k{}_{kj}\delta \bar{\Lambda}^j
+W^2\tilde{D}_j\delta \bar{\Lambda}^j
-16\pi \delta \rho_H,
\label{dif_H}
\\
\delta{\mathcal M}_i
&=&
n\tilde{A}^j{}_{i}W^{-2}(\tilde{D}_j W)(\delta W)
-n\tilde{A}^j{}_{i}W^{-1}(\tilde{D}_j\delta W)
\nonumber \\&&
+\{
n\tilde{A}^m{}_i W^{-1}(\tilde{D}^n W)
-(\tilde{D}^m \tilde{A}^n{}_{i})
\}
(\delta \tilde{\gamma}_{mn})
\nonumber \\&&
+\left(
-\tilde{A}_{i}{}^n\tilde{\gamma}^{lm}
-\frac{1}{2}\tilde{A}^{nm}\delta^l_{\ i}
+\frac{1}{2}\tilde{A}_{i}{}^l\tilde{\gamma}^{mn}\right) (\tilde{D}_l \delta\tilde{\gamma}_{mn})
\nonumber \\&&
+\frac{1-n}{n} (\tilde{D}_i \delta K) 
-n W^{-1}(\tilde{D}^m W) \delta^n_{\ i}(\delta\tilde{A}_{mn})
+\delta^n_{\ i}\tilde{\gamma}^{lm}(\tilde{D}_l \delta\tilde{A}_{mn})
\nonumber \\&&
-8\pi \delta  J_i,
\label{dif_M}
\\
\delta{\mathcal G}^i&=&
\Delta^{imn}(\delta \tilde{\gamma}_{mn})
+
\left(
-\tilde{\gamma}^{im}\tilde{\gamma}^{\ell n}
+\frac{1}{2}\tilde{\gamma}^{i\ell}\tilde{\gamma}^{m n}
\right)
(\tilde{D}_\ell \delta \tilde{\gamma}_{mn})
+\delta \bar{\Lambda}^i,
\label{dif_G}
\\
\delta{\mathcal S}&=&
\frac{\mathrm{det}(\tilde{\gamma}_{ij}) \tilde{\gamma}^{mn}}{\mathrm{det}(f_{ij})} (\delta \tilde{\gamma}_{mn}),
\label{dif_S}
\\
\delta{\mathcal{A}}&=&
-\tilde{A}^{mn}(\delta\tilde{\gamma}_{mn})
+\tilde{\gamma}^{mn}(\delta\tilde{A}_{mn}),
\label{dif_A}
\end{eqnarray}
where $\delta$ denotes the total derivative operator. 


\section*{References}
\begin{thebibliography}{99}
    \bibitem{wave} Abbott B P \etal 2016 Observation of Gravitational Waves from a Binary Black Hole Merger {\it Phys. Rev. Lett.} {\bf 116} 061102
    \bibitem{wave2} Abbott B P \etal 2016 Properties of the Binary Black Hole Merger GW150914 {\it Phys. Rev. Lett.} {\bf 116} 241102
    \bibitem{wave3} Abbott B P \etal 2016 GW151226: Observation of Gravitational Waves from a 22-Solar-Mass Binary Black Hole Coalescence {\it Phys. Rev. Lett.} {\bf 116} 241103
    \bibitem{wave4} Abbott B P \etal 2017 GW170104: Observation of a 50-Solar-Mass Binary Black Hole Coalescence at Redshift 0.2 {\it Phys. Rev. Lett.} {\bf118} 221101
    \bibitem{wave5} Abbott B P \etal 2017 GW170608: Observation of a 19 Solar-mass Binary Black Hole Coalescence {\it Astrophys. J.} {\bf 851} L35
    \bibitem{wave6} Abbott B P \etal 2017 GW170814: A Three-Detector Observation of Gravitational Waves from a Binary Black Hole Coalescence {\it Phys. Rev. Lett.} {\bf 119} 141101
    \bibitem{wave7} Abbott B P \etal 2017 GW170817: Observation of Gravitational Waves from a Binary Neutron Star Inspiral {\it Phys. Rev. Lett.} {\bf 119} 161101
    \bibitem{BSSN0} Nakamura T, Ohara K and Kojima Y 1987 General Relativistic Collapse to Black Holes and Gravitational Waves from Black Holes {\it Prog. Theor. Phys. Suppl.} {\bf 90} 1-218
    \bibitem{BSSN1} Shibata M and Nakamura T 1995 Evolution of three-dimensional gravitational waves: Harmonic slicing case {\it Phys. Rev.} D {\bf 52} 5428
    \bibitem{BSSN2} Baumgarte T W and Shapiro S L 1998 Numerical integration of Einstein's field equations {\it Phys. Rev.} D {\bf 59} 024007
    \bibitem{Z4_1} Bona C, Ledvinka T and Palenzuela C 2002 3+1 covariant suite of numerical relativity evolution systems {\it Phys. Rev.} D {\bf 66} 084013
    \bibitem{Z4_6} Alic D, Bona-Casas C, Bona C, Rezzolla L and Palenzuela C 2012 Conformal and covariant formulation of the Z4 system with constraint-violation damping {\it Phys. Rev.} D {\bf 85} 064040
    \bibitem{Z4_7} Gundlach C, Martin-Garcia J M, Calabrese G and Hinder I 2005 Constraint damping in the Z4 formulation and harmonic gauge {\it Class. Quantum Grav.} {\bf 22} 3767
    \bibitem{Z4_8} Bernuzzi S and Hilditch D 2010 Constraint violation in free evolution schemes: Comparing the BSSNOK formulation with a conformal decomposition of the Z4 formulation {\it Phys. Rev.} D {\bf 81} 084003
    \bibitem{Alic-2012-PhysRevD}
      Alic D, Bona-Casas C, Bona C, Rezzolla L and Palenzuela C 2012 Conformal and covariant formulation of the Z4 system with constraint-violation damping {\it Phys. Rev.} D {\bf 85} 064040
    \bibitem{Garfinkle-2002-PhysRevD}
      Garfinkle D 2002 Harmonic coordinate method for simulating generic singularities {\it Phys. Rev.} D {\bf 65}, 044029
    \bibitem{Pretorius-2005-PhysRevLett}
      Pretorius F 2005 Evolution of Binary Black-Hole Spacetimes {\it Phys. Rev. Lett.} {\bf 95} 121101
    \bibitem{Pretorius-2005-CQG}
      Pretorius F 2005 Numerical relativity using a generalized harmonic decomposition {\it Class. Quantum Grav.} {\bf 22} 425
    \bibitem{spectral1} Tichy W 2006 Black hole evolution with the BSSN system by pseudospectral methods {\it Phys. Rev.} D {\bf 74} 084005
    \bibitem{spectral2} Scheel M A, Pfeiffer H P, Lindblom L, Kidder L E, Rinne O and Teukolsky S A 2006 Solving Einstein{’s} equations with dual coordinate frames {\it Phys. Rev.} D {\bf 74} 104006
    \bibitem{spectral3} Meringolo C, Servidio S and Veltri P 2021 A spectral method algorithm for numerical simulations of gravitational fields {\it Class. Quantum Grav.} {\bf 38} 075027
    \bibitem{spectral4} Rashti A, Fabbri F M, Br\"{u}gmann B, Chaurasia S V, Dietrich T, Ujevic M and Tichy W 2022 New pseudospectral code for the construction of initial data {\it Phys. Rev.} D {\bf 105} 104027
    \bibitem{spectral5} Imbrogno M, Meringolo C and Servidio S 2021 Strong Interactions in the Three Black Holes Problem arXive:2108.01392
    \bibitem{improvedBSSN2} Tichy W 2009 Long term black hole evolution with the BSSN system by pseudospectral methods {\it Phys. Rev.} D {\bf 80} 104034
    \bibitem{alcubierre2008introduction}
      Alcubierre M 2008 Introduction to $3+1$ numerical relativity (Oxford University Press)
    \bibitem{baumgarte2010numerical}
      Baumgarte T W and Shapiro S L 2010 Numerical Relativity: Solving Einstein{'s} Equations on the Computer (Cambridge University Press)
    \bibitem{Gourgoulhon-2012}
      Gourgoulhon E 2012 $3+1$ Formalism in General Relativity (Springer, New York)
    \bibitem{instabilityBSSN1} Montero P J and Cordero-Carri\`{o}n 2012 BSSN equations in spherical coordinates without regularization: Vacuum and nonvacuum spherically symmetric spacetimes {\it Phys. Rev.} D {\bf 85} 124037
    \bibitem{instabilityBSSN2} Meringolo C and Servidio S 2021 Aliasing instabilities in the numerical evolution of the Einstein field equations {\it Gen. Relativ. Gravit.} {\bf 53} 95 
    \bibitem{cBSSN1} Brown J D 2009 Covariant formulations of Baumgarte, Shapiro, Shibata and Nakamura and the standard gauge {\it Phys. Rev.} D {\bf 79} 104029
     \bibitem{Alcubierre-2011-GRG}
      Alcubierre M and Mendez M D 2011 Formulations of the $3+1$ evolution equations in curvilinear coordinates {\it Gen. Relativ. Gravit.}
      {\bf 43} 2769
    \bibitem{cBSSN_2012} Brown J D, Diener P, Field S E, Hesthaven J S, Herrmann F,  Mrou\`{e} A H, Sarbach O, Schnetter E, Tiglio M and Wagman M 2012 Numerical simulations with a first-order BSSN formulation of Einstein's field equations {\it Phys. Rev.} D {\bf 85} 084004
    \bibitem{cBSSN2} Baumgarte T W, Montero P J, Cordero-Carri\'{o}n I and M\"{u}ller E 2013 Numerical relativity in spherical polar coordinates: Evolution calculations with the BSSN formulation {\it Phys. Rev.} D {\bf 87} 044026
    \bibitem{cBSSN3} Baumgarte T W, Montero P J and M\"{u}ller E 2015 Numerical relativity in spherical polar coordinates: Off-center simulations {\it Phys. Rev.} D {\bf 91} 064035
    \bibitem{Akbarian-2015-PhysRevD}
      Akbarian A and Choptuik M W 2015 Black hole critical behavior with the generalized BSSN formulation
      {\it Phys. Rev.} D {\bf 92} 084037
    \bibitem{cBSSN4} Ruchlin I, Etienne Z B and Baumgarte T W 2018 $\mathrm{SENR}/\mathrm{NRPy}+$: Numerical relativity in singular curvilinear coordinate systems {\it Phys. Rev.} D {\bf 97} 064036
    \bibitem{cBSSN5} Mewes V, Zlochower Y, Camanpanelli M, Ruchlin I, Etienne Z B and Baumgarte T W 2018 Numerical relativity in spherical coordinates with the Einstein Toolkit {\it Phys. Rev.} D {\bf 97} 084059
    \bibitem{cBBSN_2020} Torsello F, Kocic M, H\"og\r{a}s M and M\"ortsell E 2020 Covariant BSSN formulation in bimetric relativity {\it Class. Quant. Grav.} {\bf 37} 079501
     \bibitem{PRD104_084065}
       Alcoforado M A, Aranha R F, Barreto W O and de Oliveira H P
      2021 New numerical framework for the generalized Baumgarte-Shapiro-Shibata-Nakamura formulation: The vacuum case for spherical symmetry {\it Phys. Rev.} D {\bf 104} 084065
    \bibitem{covariantZ4_1} Sanchis-Gual N, Montero P J,  Font J A, M\"{u}ller E and Baumgarte T W 2014 Fully covariant and conformal formulation of the Z4 system in a reference-metric approach: Comparison with the BSSN formulation in spherical symmetry {\it Phys. Rev.} D {\bf 89} 104033
    \bibitem{covariantZ4_2} Alic D, Kastaun W and Rezzolla L 2013 Constraint damping of the conformal and covariant formulation of the Z4 system in simulations of binary neutron stars {\it Phys. Rev.} D {\bf 88} 064049
    \bibitem{adjusted1} Yoneda G and Shinkai H 2001 Hyperbolic formulations and numerical relativity: II. asymptotically constrained systems of Einstein equations {\it Class. Quantum Grav.} {\bf 18} 441-462
    \bibitem{adjusted2} Yoneda G and Shinkai H 2001 Constraint propagation in the family of ADM systems {\it Phys. Rev.} D {\bf 63} 124019
    \bibitem{adjusted3} Yoneda G and Shinkai H 2002 Advantages of a modified ADM formulation: Constraint propagation analysis of the Baumgarte-Shapiro-Shibata-Nakamura system {\it Phys. Rev.} D {\bf 66} 124003
    \bibitem{adjusted4} Shinkai H and Yoneda G 2002 Adjusted ADM systems and their expected stability properties: constraint propagation analysis in Schwarzschild spacetime {\it Class. Quantum Grav.} {\bf 19} 1027-1049
     \bibitem{improvedBSSN1} Yo H-J, Baumgarte T W and Shapiro S L 2002 Improved numerical stability of stationary black hole evolution calculations {\it Phys. Rev.} D {\bf 66} 084026
      \bibitem{adjusted6} Tsuchiya T, Yoneda G and Shinkai H 2012 Constraint propagation of ${C}^{2}$-adjusted formulation. II. Another recipe for robust Baumgarte-Shapiro-Shibata-Nakamura evolution system {\it Phys. Rev.} D {\bf  85} 044018
    \bibitem{Frittelli-1997-PhyRevD}
      Fritteri S 1997  Note on the propagation of the constraints in standard 3+1 general relativity
      {\it Phys. Rev.} D
      {\bf 55}
      5992--5996
    \bibitem{adjusted5} Tsuchiya T, Yoneda G and Shinkai H 2011 Constraint propagation of ${C}^{2}$-adjusted formulation: Another recipe for robust ADM evolution system {\it Phys. Rev.} D {\bf 83} 064032
    \bibitem{adjusted8} Urakawa R, Tsuchiya T and Yoneda G 2019 Analyzing time evolution of constraint equations of Einstein's equation {\it JSIAM Lett.} {\bf 11} 21-24
\end{thebibliography}
\end{document}